\documentclass[twocolumn,pra,showpacs,preprintnumbers,amsmath,amssymb]{revtex4}

\usepackage{graphicx}

\begin{document}
\title{Homodyne detection and parametric down-conversion: a classical 
approach applied to proposed ``loophole-free'' Bell tests}
\author{Caroline H Thompson}
  \email{ch.thompson1@virgin.net}
\affiliation{
11 Parc Ffos, Ffos-y-Ffin, Aberaeron, SA46 0HS, U.K.
}

\date{\today}

\begin{abstract}

A classical model is presented for the features of parametric 
down-conversion and homodyne detection relevant to recent proposed 
``loophole-free'' Bell tests. The Bell tests themselves are uncontroversial: 
there are no obvious loopholes that might cause bias and hence, if the world 
does, after all, obey local realism, no violation of a Bell inequality will 
be observed. Interest centres around the question of whether or not the 
proposed criterion for ``non-classical'' light is valid. If cit is not, then 
the experiments will fail in their initial concept, since both quantum 
theorists and local realists will agree that we are seeing a purely 
classical effect. The Bell test, though, is not the only criterion by which 
the quantum-mechanical and local realist models can be judged. If the 
experiments are extended by including a range of parameter values and by 
analysing, in addition to the proposed digitised voltage differences, the 
raw voltages, the models can be compared in their overall performance and 
plausibility.
\end{abstract}

\maketitle

\section{Introduction}
No test of Bell's inequalities \cite{Bell:1964, Bell:1971} to date has been free 
of ``loopholes''. This 
means that, despite the high levels of statistical significance frequently 
achieved, violations of the inequalities could be the effects of 
experimental bias of one kind or another, not evidence for the presence of 
quantum entanglement. Recent proposed experiments by Garc\'{\i}a-Patr\'{o}n 
S\'{a}nchez \cite{Sanchez:2004}\textit{ et al.} and Nha and Carmichael \cite{Nha:2004} show 
promise of being genuinely free from such problems. If the world in fact 
obeys local realism, they should \textit{not}, therefore, infringe any Bell inequality.

\begin{figure}[htbp]
\centerline{\includegraphics[width=2.6in,height=2.6in]{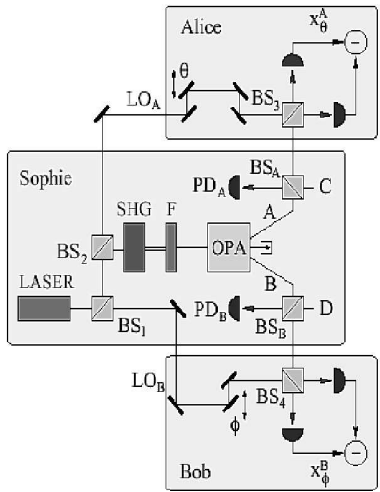}}
\caption{Proposed experimental set-up. 
In the current text, phase shifts $\theta $ and $\phi $ are renamed $\theta 
_{A}$ and $\theta _{B}$. (\textit{Reproduced with permission from S\'{a}nchez et al., 
Phys. Rev. Lett. }\textbf{\textit{93}}\textit{, 130409 (2004)})
}
\label{fig1}
\end{figure}

The current article presents a classical (local realist) model that should 
be able, once all relevant details are known, to explain the results. It 
uses standard classical theory for homodyne detection, but for the behaviour 
of the ``Optical Parametric Amplification'' (``Parametric Down-Conversion'') 
source introduces new ideas.

As far as the ``loophole-free'' status of the proposed experiments is 
concerned, there would appear to be no problem. A difficulty that seems 
likely to arise, though, is that theorists may not agree that the test beams 
used were in fact ``non-classical'', so the failure to infringe a Bell 
inequality will not in itself be interpreted as showing a failure of quantum 
mechanics \cite{prediction}. The criterion to be used to establish the 
non-classical nature of the light is the observation of negative values of 
the Wigner density, and there is reason to think that, even if the standard 
method of estimation seems to show that these are achieved, this may not in 
fact be so. 

In the quantum mechanical theory discussed in the proposals and in other 
recent papers \cite{Lvovsky:2001,Wenger:2004}, the presence of 
negative Wigner densities is seen almost as a corollary of the observation of 
double peaks in the distribution of the phase-averaged voltage differences 
recorded by the homodyne detectors. Such double peaks, however, can readily 
be shown to occur naturally, under purely classical assumptions, so long as 
noise levels are low. They are a consequence of the fact that a sine 
function spends proportionately more time near the extremes than around the 
central, zero, value. It would seem that the theory that leads from their 
observation to the deduction of non-classicality may be in 
error. This, though, is a problem that is of no direct relevance to the 
classical model under consideration. Far from being, as suggested by S\'{a}nchez  
and others, the ``hidden variable'' needed, Wigner density plays no part whatsoever. 

Regardless of the outcome of the Bell tests, and whether or not the light is 
declared to be non-classical, there are features of the experiments that can 
usefully be exploited to compare the strengths of quantum mechanics versus 
(updated) classical theory as viable models. The two theories approach the 
situation from very different angles. Classical theory traces the causal 
relationships between phenomena, starting with the simplest assumption and 
building in random factors later where necessary. Quantum mechanics starts 
with models of complete ensembles, all random factors included. This, it is 
argued, is inappropriate, since two features of the proposed experiments 
demand that we consider the behaviour of individual events, not whole 
ensembles: the process of homodyne detection itself, and the Bell test.

\section{The proposed experiments}

The experimental set-up proposed by S\'{a}nchez \textit{et al.} is shown 
in Fig.~1, that of Nha and Carmichael being similar. In the words of the 
S\'{a}nchez \textit{et al.} proposal:

\begin{quotation}
\noindent
The source (controlled by Sophie) is based on a master laser beam, which 
is converted into second harmonic in a nonlinear crystal (SHG). After 
spectral filtering (F), the second harmonic beam pumps an optical parametric 
amplifier (OPA) which generates two-mode squeezed vacuum in modes A and B. 
Single photons are conditionally subtracted from modes A and B with the use 
of the beam splitters BS$_{A}$ and BS$_{B}$ and single-photon detectors 
PD$_{A}$ and PD$_{B}$. Alice (Bob) measures a quadrature of mode A (B) using 
a balanced homodyne detector that consists of a balanced beam splitter 
BS$_{3}$ (BS$_{4 })$ and a pair of highly-efficient photodiodes. The local 
oscillators LO$_{A}$ and LO$_{B}$ are extracted from the laser beam by means 
of two additional beam splitters BS$_{1}$ and BS$_{2}$.
\end{quotation}

The classical description, working from the same figure, is just a little 
different. Quantum theoretical terms such as ``squeezed vacuum'' and are not 
used since they are not appropriate to the model and would cause confusion. 
The description might run as follows:

The master laser beam (which is, incidentally, pulsed) is frequency-doubled 
in the crystal SHG. After filtering to remove the original frequency, the 
beam is used to pump the crystal OPA, which outputs pairs of classical wave 
pulses at half the input frequency, i.e.~at the original laser frequency. 
The selection of pairs for analysis is done by splitting each output at an 
unbalanced beamsplitter (BS$_{A}$ or BS$_{B})$, the smaller parts going to 
sensitive detectors PD$_{A}$ or PD$_{B}$. Only if there are detections at 
both PD$_{A}$ and PD$_{B}$ is the corresponding homodyne detection included 
in the analysis. The larger parts proceed to balanced homodyne detectors, 
i.e.~ones in which the intensities of local oscillator and test inputs are 
approximately equal. The source of the local oscillators LO$_{A}$ and 
LO$_{B}$ is the same laser that stimulated, after frequency doubling, the 
production of the test beams. 

\section{Homodyne detection}
In (balanced) homodyne detection, the test beam is mixed at a beamsplitter 
with a local oscillator beam of the same frequency and the two outputs sent 
to photodetectors that produce voltage readings for every input pulse. In 
the proposed ``loophole-free'' Bell test the difference between the two 
voltages will be converted into a digital signal by counting all positive 
values as +1, all negative as --1.

\begin{figure}[htbp]
\centerline{\includegraphics[width=1.5in,height=1.5in]{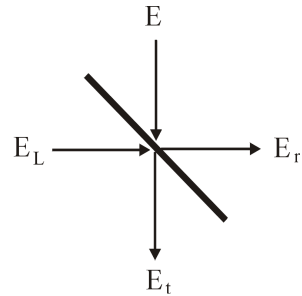}}
\caption{Inputs and outputs at the beamsplitter in a homodyne detector.
E$_{L }$ is the local oscillator beam, E the test beam, E$_{t }$ and E$_{r}$  
the transmitted and reflected beams respectively.}
\label{fig2}
\end{figure}

Assuming the inputs are all classical waves of the same frequency and there 
are no losses, it can be shown (see below) that the difference between the 
intensities of the two output beams is proportional to the product of the 
two input intensities multiplied by $\sin \theta$, where $\theta$ is the phase 
difference between the test beam and local oscillator. If voltages are 
proportional to intensities then it follows that the voltage difference will 
be proportional to $\sin \theta $. When digitised, this transforms to a step function, 
taking the value $-1$ for $-\pi < \theta < 0$ and $+1$ for $0 < \theta < \pi $.  
(The function is not well defined for integral multiples of $\pi$.)

\subsection{Classical derivation of the predicted voltage difference}
Assume the test and local oscillator signals have the same frequency, 
\textit{$\omega $}, the time-dependent part of the test signal being modelled by $e^{i\phi }$, 
where (ignoring a constant phase offset \cite{phase offset}) \textit{$\phi =\omega $t} is the phase angle, 
and the local oscillator phase and test beam phases differ by \textit{$\theta $}. [Note that 
although complex notation is used here, only the real part has meaning: this 
is an ordinary wave equation, not a quantum-mechanical ``wave function''. To 
allay any doubts on this score, the derivation is partially repeated with no 
complex notation in the Appendix.]

\begin{figure}[htbp]
\centerline{\includegraphics[width=1.8in,height=1.8in]{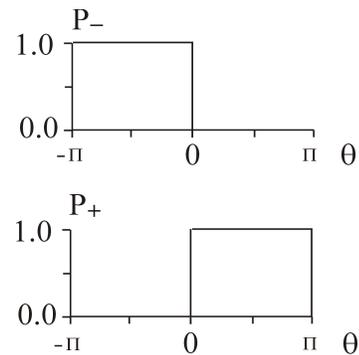}}
\caption{Probabilities of `+' and `--' outcomes versus phase difference, 
using digitised difference voltages from a perfect, noise-free, balanced 
homodyne detector.}
\label{fig3}
\end{figure}

Let the electric fields of the test signal, local oscillator and reflected and transmitted signals 
from the beamsplitter have amplitudes $E$, $E_{L}$, $E_{r}$ and $E_{t}$ respectively, as 
shown in Fig.~\ref{fig2}. Then, after allowance for phase delays of $\pi/2$ at each 
reflection and assuming no losses, we have
\begin{equation}
\label{eq1}
E_r = \frac{1}{\sqrt 2} (Ee^{i(\phi +\pi /2)}+E_L e^{i(\phi +\theta )})
\end{equation}
and
\begin{equation}
\label{eq2}
E_t = \frac{1}{\sqrt 2} (Ee^{i\phi }+E_L e^{i(\phi +\theta +\pi /2)}).
\end{equation}
The intensity of the reflected beam is therefore
\begin{eqnarray}
\label{eq3}
E_r E_r^\ast & = & \frac{1}{2} (Ee^{i(\phi +\pi /2)}+E_L e^{i(\phi +\theta )})(Ee^{-i(\phi 
+\pi /2)}
\nonumber \\
             & + & E_L e^{-i(\phi +\theta )})
\nonumber \\
             & = & \frac{1}{2}(E^2+E_L ^2+EE_L e^{i(\pi /2-\theta )}+EE_L e^{-i(\pi /2-\theta )})
\nonumber \\
             & = & \frac{1}{2} (E^2+E_L ^2+2EE_L \cos (\pi /2-\theta )
\nonumber \\
             & = & \frac{1}{2} (E^2+E_L ^2+2EE_L \sin \theta ).
\end{eqnarray}
Similarly, it can be shown that the intensity of the transmitted beam is
\begin{equation}
\label{eq4}
E_t E_t^\ast = \frac{1}{2} (E^2+E_L ^2-2EE_L \sin \theta ).
\end{equation}
If the voltages registered by the photodetectors are proportional to the 
intensities, it follows that the difference in voltage is proportional to 
$2EE_L \sin \theta$. When digitised, this translates to the 
step function mentioned above. The probabilities for the two possible 
outcomes are, as shown in Fig.~\ref{fig3},

\begin{subequations}
\label{eqs5}
\begin{equation}
p_{-} = \left\{
\begin{array}{ll}
     1 & \mbox{for $-\pi < \theta < 0$} \\
     0 & \mbox{for $0 < \theta < \pi$}
\end{array}
\right.
\label{subeq5a}
\end{equation}
\\ \mbox{and} \\
\begin{equation}
p_{+} = \left\{
\begin{array}{ll}
     0 & \mbox{for $-\pi < \theta < 0$} \\
     1 & \mbox{for $0 < \theta < \pi$}
\end{array}
\right.
\label{subeq5b}
\end{equation}
\end{subequations}

Note that the probabilities are undefined for integral multiples of $\pi$.  
In practice it would be reasonable to assume that, due to the presence of noise,
 all the values were 0.5, but for the present purposes the integral values will
simply be ignored.

\section{Fresh ideas on parametric down-conversion}
If the frequencies and phases of both test beams and both local oscillators 
were all identical apart from the applied phase shifts, the experiment would 
be expected to produce step function relationships between counts and 
applied shifts both for the individual (singles) counts and for the 
coincidences. 

\begin{figure}[htbp]
\centerline{\includegraphics[width=2.7in,height=1.8in]{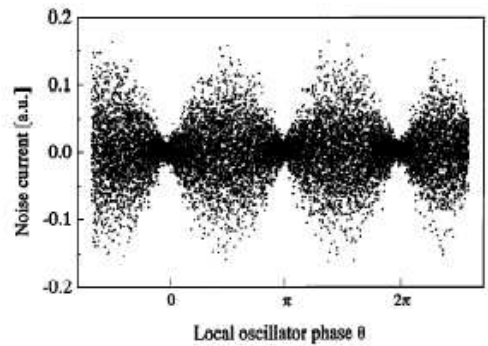}}
\caption{A typical scatter of ``noise current'' (related to voltage difference) for 
varying local oscillator phase. (\textit{Reproduced from S. Schiller et al., 
Phys. Rev. Lett. 87 (14), 2933 (1996)}. Permission requested, June 12, 2005.)}
\label{fig4}
\end{figure}

It may safely be assumed that this is not what is observed. It would have 
shown up in the preliminary trials on the singles counts (see ref.~\cite{Wenger:2004}), 
which would have followed something suggestive of the basic predicted step 
function as the local oscillator phase shift was varied. What is observed in 
practice is more likely to be similar to the results obtained by Schiller 
\textit{et al.} \cite{Schiller:1996}. Their Fig.~2a, reproduced here as Fig.~\ref{fig4}, shows a 
distribution of photocurrents that is clustered around zero, for $\theta $ 
taking integer multiples of $\pi $, but is scattered fairly equally among 
positive and negative values in between. 

When digitised, the distribution would reduce to two straight horizontal 
lines, showing that for each choice of $\theta $ there is an equal chance of 
a `$+$' or a `$-$' result. As in any other Bell test setup, though, the absence 
of variations in the singles counts does not necessarily mean there is no variation in 
the coincidence rates.  However, as explained in the next section, the 
predicted coincidence curves are not the zig-zag ones of standard classical 
theory. These would be expected if we had \textit{full}
``rotational invariance'' \cite{rotational invariance}.  If the ideas presented here are 
correct, we have instead, in the language of an article by the 
present author \cite{Thompson:1999}, only \textit{ binary} rotational invariance.  
Schiller's scatter of photocurrent differences is 
seen as evidence that the relative phase can (under perfect conditions) take 
just one of two values, 0 or $\pi$.  The scatter is formed from a superposition 
of two sets of points, corresponding to two sine curves that are out of phase, 
together with a considerable amount of noise.

It is suggested that the two ``phase sets'' arise from the way in which the 
pulses are produced, which involves, after the frequency doubling, the 
\textit{degenerate} case of parametric down-conversion, the latter producing pulses that are 
(in contrast to the quantum-mechanical assumption of conjugate frequencies) of 
\textit{exactly equal frequency}. Consider an initial pump laser of frequency $\omega $. In the proposed 
experiment, this will be doubled in the crystal SHG to 2$\omega $ then 
halved in OPA back to $\omega $. At the frequency doubling stage, one laser 
input wave peak gives rise to two output ones. Assuming that there are 
causal mechanisms involved, it seems inevitable that every other wave peak 
of the output will be exactly in phase with the input. When we now use this 
input to produce a down-conversion, the outputs will be in phase either with 
the even or with the odd peaks of the input, which will make them either in 
phase or exactly out of phase with the original laser.

We thus have two classes of output, differing in phase by $\pi $. If we 
define the random variable $\alpha $ to be 0 
for one class, $\pi $ for the other, this will clearly be an important 
``hidden variable'' of the system.

The theory of (degenerate) parametric down-conversion assumed here differs 
in several respects from the accepted quantum-mechanical one, and also from 
Stochastic Electrodynamics. No attempt is made to give a full explanation of 
the physics involved. The theory is rather of the nature of an empirical 
result, effectively forced on us by a number of different experiments in 
quantum optics (to be discussed in later papers). Accepted theory says (if, 
indeed, it allows at all for the existence of ``photons'' with definite frequencies
 \cite{Kwiat:1991}) that though the sum of the frequencies of the two 
outputs is equal to that of the pump laser, even in the degenerate case the 
two will differ slightly \cite{Tittel:1998}. The exact frequency of one will 
be $\frac{1}{2}(\omega _{0}+\delta \omega)$, that of the other 
$\frac{1}{2}(\omega _{0}-\delta \omega)$, where $\omega _{0}$ is the pump frequency. 
If this is the case in the proposed experiment, though, it will severely reduce 
the visibility of any coincidence curve observed when the experimental beam is 
mixed back with the source laser in the homodyne detector.

The preliminary experiments (see ref.~\cite{Wenger:2004}) using just one output beam may 
already be sufficient to show that the interference is stronger than would 
be the case if there were any difference between local oscillator and test 
beam frequencies. It is known that the source laser has quite a broad band 
width, i.e.~that $\omega _{0}$ is not constant. Though it is likely that 
it is only part of the pump spectrum that induces a 
down-conversion, so that the band width of the test beam may be considerably 
narrower than that of the pump, it too is non zero. It follows that 
agreement of frequency between this and the test beam must be because we are 
always dealing, in the degenerate case, with \textit{exact} frequency doubling and 
halving.

\section{A classical model of the proposed Bell test}
In the proposed Bell test of S\'{a}nchez \textit{et al.}, positive voltage 
differences will be treated as +1, negative as --1. Applying this version of 
homodyne detection 
to both branches of the experiment, the CHSH test ($ -2 \leq S \leq 2$) will 
then be applied to the coincidence counts. Under quantum theory it is 
expected that, so long as ``non-classical'' beams are employed, the test 
will be violated. However, since there are no obvious loopholes in the 
actual Bell test (see later), local realism should win: the test should 
\textit{not} be violated. In the classical view, this prediction is unrelated to any 
supposed non-classical nature of the light.

\subsection{The basic local realist model}
If we take the simplest possible case, in which to all intents and purposes 
all the frequencies involved are the same, the hidden variable in the local 
realist model is clearly going to be the phase difference ($\alpha = 0$ or $\pi$) 
between the test signal and the local oscillator.  If high visibility 
coincidence curves are seen, it must be because the values of $\alpha$ are identical for
the A and B beams.  Assuming no noise, the basic model is easily written down.

From equation (\ref{subeq5a}), the probability of a $-1$ outcome on side A is

\begin{equation}
\label{eq6}
p_{-} (\theta _{A}, \alpha )= \left\{
\begin{array}{ll}
    1 & \mbox{for $-\pi  <  \theta _{A} - \alpha  < 0$} \\
    0 & \mbox{for $ 0 < \theta _{A} - \alpha < \pi $},
\end{array}
\right.
\end{equation}

\noindent
where $\theta _{A}$ is the phase shift applied to the local oscillator A, 
$\alpha$ is the hidden variable and all angles are reduced modulo $2\pi$. 
Similarly, the probability of a +1 outcome is

\begin{equation}
\label{eq7}
p_{+} (\theta _{A}, \alpha )= \left\{
\begin{array}{ll}
    0 & \mbox{for $-\pi  <  \theta _{A} - \alpha  < 0$} \\
    1 & \mbox{for $ 0 < \theta _{A} - \alpha < \pi $},
\end{array}
\right.
\end{equation}

Assuming equal probability $\frac{1}{2}$ for each of the two possible values of 
$\alpha$ \cite{equal probability}, the standard ``local realist'' assumption that independent 
probabilities can be multiplied to give coincidence ones leads to a 
predicted coincidence rate of
\begin{eqnarray}
\label{eq8}
P_{++} (\theta _A ,\theta _B)& = & \frac{1}{2} p_+ (\theta _A ,0)p_+ (\theta _B 
,0)
\nonumber \\
                        & + & \frac{1}{2} p_+ (\theta _A ,\pi )p_+ (\theta _B ,\pi ),
\end{eqnarray}
with similar expressions for $P_{+-}$, $P_{-+ }$ and $P_{- -}$. 

\noindent The result for $\theta _{A}=\pi/2$, for example, is
\begin{equation}
\label{eq9}
P_{++} (\pi /2, \theta _{B}) = \left\{
\begin{array}{ll}
     0 &  \mbox{for $-\pi  <  \theta _{B} <  0$}\\
   1/2 & \mbox{for $0 < \theta _{B} < \pi$}.
\end{array}
\right.
\end{equation}
For $\theta _{A}$ = --$\pi $/2 it is
\begin{equation}
\label{eq10}
P_{++} (-\pi /2, \theta _{B}) = \left\{
\begin{array}{ll}
     1/2 &  \mbox{for $-\pi  <  \theta _{B} <  0$}\\
       0 & \mbox{for $0 < \theta _{B} < \pi$}.
\end{array}
\right.
\end{equation}

\begin{figure}[htbp]
\centerline{\includegraphics[width=1.8in,height=2.7in]{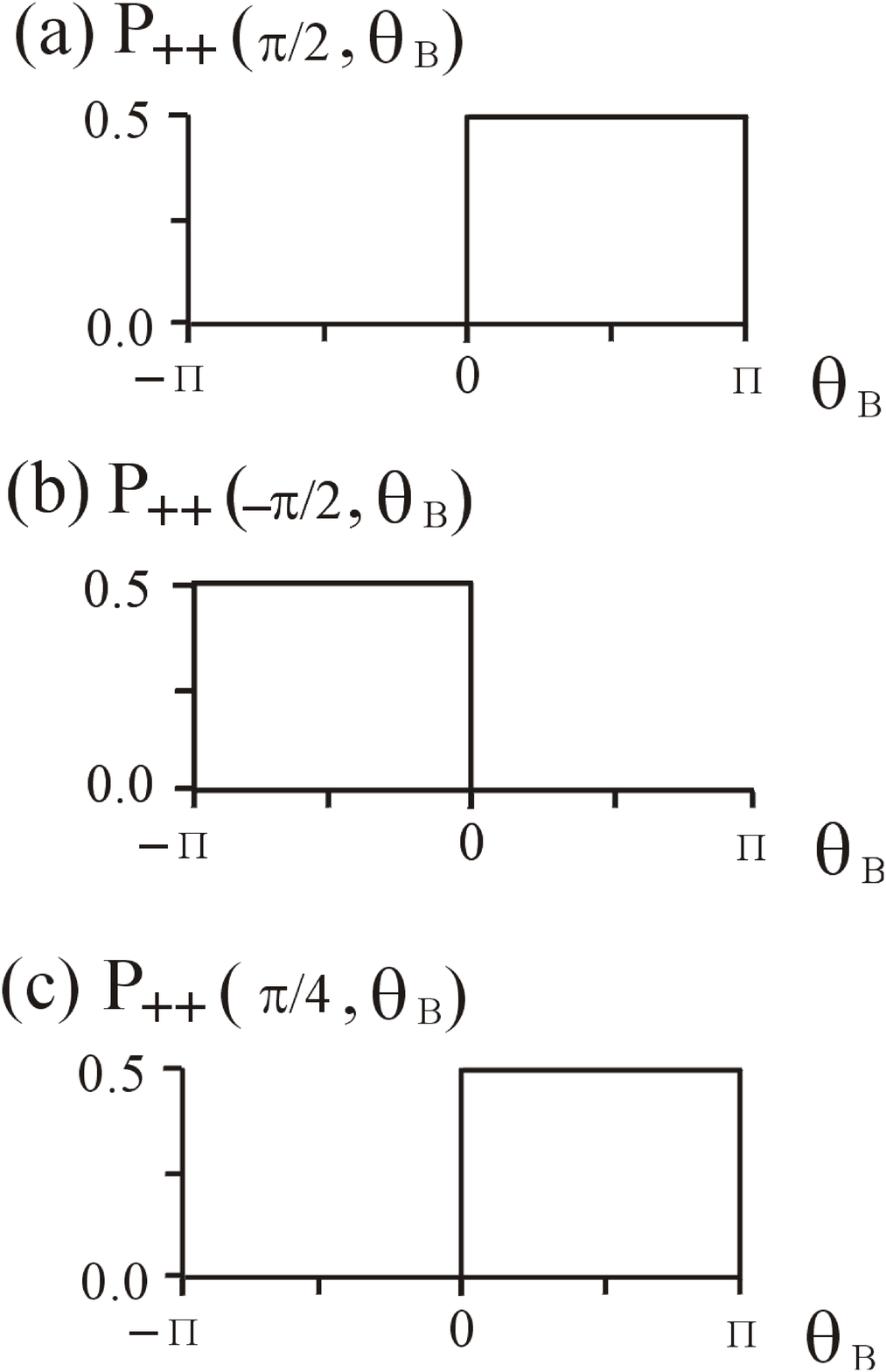}}

\caption{Predicted coincidence curves for the ideal experiment.  
(a) and (b) illustrate the settings most likely to be chosen in practice, 
giving the strongest correlations. $\theta _{A}$ is fixed at $\pi/2$ or 
$-\pi/2$ while $\theta _{B }$ varies. In theory, any value of $\theta 
_{A}$ between 0 and $\pi$ would give the same curve as (a), any between 
$-\pi$ and 0 the same as (b). An example is shown in (c), where $\theta 
_{A }$ is $\pi/4$ but the curve is identical to (a). We do not have 
rotational invariance: the curve is not a function of $\theta _{B }-\theta _{A}$.
}
\label{fig5}
\end{figure}

Note that, as illustrated in Fig.~\ref{fig5}, the coincidence probabilities \textit{cannot}, in this 
basic model, be expressed as functions of the difference in detector 
settings, $\theta _{B}- \theta _{A}$. This failure, marking a significant deviation 
from the quantum mechanical prediction, is an inevitable consequence of the fact that 
we have (as mentioned earlier) only binary, not full, rotational invariance.

\subsection{Fine-tuning the model}
Many practical considerations mean that the final local realist prediction 
will probably not look much like the above step function. It may not even be 
quite periodic. The main logical difference is that, despite all that has 
been said so far, the actual variable that is set for the local oscillators 
is not directly the phase shift but the path length, and, since the 
frequency is likely to vary slightly from one signal to the next (though 
always keeping the same as that of the pump laser), the actual phase 
difference between test and local oscillator beams will depend on the path 
length difference \textit{and} on the frequency. In a complete model, therefore, the 
important parameters will be path length and frequency, with phase derived 
from these.

If frequency variations are sufficiently large, the situation may approach 
one of rotational invariance (RI), but it seems on the face of it unlikely 
that this can be achieved without loss of correlation. If we do have RI, the 
model becomes the standard realist one in which the predicted quantum 
correlation varies linearly with difference in phase settings, but it is 
more likely that what will be found is curves that are \textit{not} independent of the 
choice of individual phase setting. They will be basically the predicted 
step functions but converted to curves as the result of the addition of 
noise.

\begin{figure}[htbp]
\centerline{\includegraphics[width=2.0in,height=1.0in]{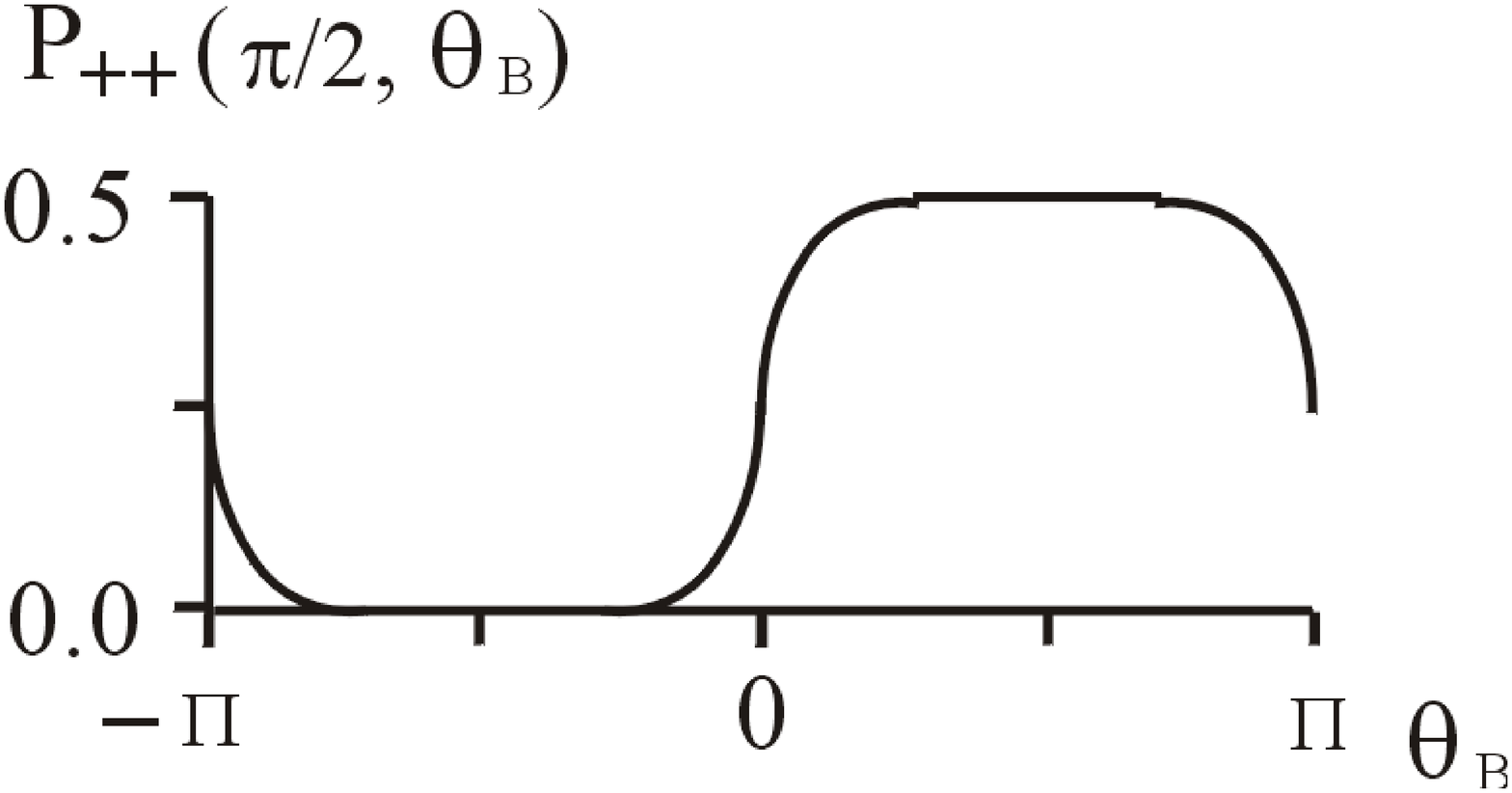}}
\caption{Likely appearance of coincidence curves in a real experiment with 
moderate noise.}
\label{fig6}
\end{figure}

It is essential to know the actual experimental conditions. Several relevant 
factors can be discovered by careful analysis of the variations in the raw 
voltages in each homodyne detection system. If noise is low, the presence of 
the two phase sets, and whether or not they are equally represented, should 
become apparent. All this complexity, though, has no bearing on the basic 
fact of the existence of a hidden variable model and the consequent 
prediction that the CHSH Bell test will not be violated.

\subsection{The role of the ``event-ready'' detectors}
In the quantum-mechanical theory, the expectation of violation of the Bell 
test all hinges on the production of ``non-classical'' light. The light 
directly output from the crystal OPA is assumed to be Gaussian, i.e.~it 
takes the form of pulses of light that have a Gaussian intensity profile and 
also, as a result of Fourier theory, a Gaussian spectrum. When this is 
passed through an unbalanced beamsplitter (BS$_{A}$ or BS$_{B}$) and a 
``single photon'' detected by an ``event-ready'' detector, the theory says 
that the subtraction of one photon leaves the remaining beam 
``non-Gaussian''. Although there is mention here of single photons, the 
theory is concerned with the ensemble properties of the complete beams, not 
with the individual properties of its constituent pulses.

In the local realist (classical) model, the shapes of the spectra are not
relevant except insofar as a narrow band width is desirable for the 
demonstration of dramatic correlations. The event-ready detectors play, 
instead, the important role of selecting for analysis only the strongest 
down-converted output pairs, it being assumed that the properties of the 
transmitted and reflected light at the unbalanced beamsplitters are 
identical apart from their amplitudes. It is likely that 
those detected signals that are coincident with each other will be genuine 
``degenerate'' ones, i.e.~of exactly equal frequency, quasi-resonant with 
the pump laser. The unbalanced beamsplitters and the detectors PD$_{A}$ 
and PD$_{B}$ need to be set so that 
the intensity of the detected part is sufficient to be above the minimum for 
detection but low enough to ensure that all but the strongest pulses are 
ignored.

In neither theory are the event-ready detectors really needed in their 
``Bell test'' role of ensuring a fair test (see below), since the homodyne 
detectors are almost 100{\%} efficient.

\section{Validity of the proposed Bell test}
Coincidences between the digitised voltage differences will be used in the 
CHSH Bell test \cite{Clauser:1969, Thompson:1996}, but avoiding the ``post-selection'' 
that has, since 
Aspect's second experiment \cite{Aspect:1982}, become customary.
The S\'{a}nchez \textit{et al.} proposal is to use ``event-ready'' detectors, 
as recommended by Bell himself for use in real experiments \cite{Clauser:1978}.
None of the usual loopholes \cite{Thompson:2003} are expected to be applicable:

\begin{enumerate}
\item With the use of the event-ready detectors, non-detections are of little concern. 
The detectors (PD$_{A}$ and PD$_{B}$ in Fig.~\ref{fig1}) act to define the sample to be analysed, 
and the fact that they do so quite independently of whether or not any member of the sample 
is then also detected in coincidence at the homodyne detectors ensures that no bias is 
introduced here. The estimate of ``quantum correlation'' \cite{quantum correlation} 
to be used in calculating the 
CHSH test statistic is $E = (N_{++AB} + N_{--AB} - N_{+-AB} - N_{-+AB}) / N_{AB}$, where 
the $N$'s are coincidence 
counts and the subscripts are self-explanatory. This contrasts with the usual method, in 
which the denominator used is not $N_{AB}$ but the sum of observed coincidences, 
$N_{++AB} + N_{--AB} + N_{+-AB} + N_{-+AB}$. The use of the latter can readily be shown 
to introduce bias unless it can be assumed that the sample of detected pairs is a fair one. 
That such an assumption fails in some plausible local realist models has been known 
since 1970 or earlier \cite{Pearle:1970, Thompson:1996}. 
\item There is no possibility of synchronisation problems, since a pulsed source is used.
\item No ``accidentals'' will be subtracted. 
\item The ``locality'' loophole can be closed by using long paths and a random system for 
choosing the ``detector settings'' (local oscillator phase shifts) during the propagation of 
the signals.
\end{enumerate}

The system is almost certainly not going to be ``rotationally invariant'' 
(not all phase differences will be equally likely), but this will not 
invalidate the Bell test. It may, however, be important in another way. It 
is likely that high visibilities will be observed in the coincidence curves 
(i.e.~high values of (max -- min)/(max + min) in plots of coincidence rate 
against difference in phase shift), leading to the impression that the Bell 
test ought to be violated. These visibilities, though, will depend on 
the absolute values of the individual settings. High ones will be balanced by low, 
with the net effect that violation does not in fact happen.

\section{Validity and significance of negative estimates for Wigner densities}
\begin{figure}[htbp]
\centerline{\includegraphics[width=2.7in,height=2.00in]{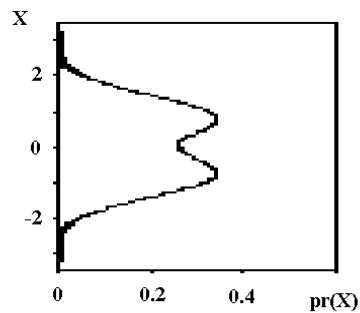}}
\caption{Observed distribution of voltage differences, X, using homodyne 
detection in an experiment similar to the proposed Bell test and averaging 
over a large number of applied phase shifts. (\textit{Based on Fig.~4a of 
A.~I.~Lvovsky et al., Phys. Rev. Lett, 87, 050402 (2001).})}
\label{fig7}
\end{figure}

Part of the evidence that is put forward as indicating that negative Wigner 
densities are likely to be obtained consists in the observation that, when 
$\theta $ is varied randomly, the distribution of observed voltage 
differences shows a double peak (see Fig.~\ref{fig7}). There is a tendency to observe 
roughly equal numbers of + and -- results but relatively few near zero. The 
fact that the relationship depends on the sine of $\theta$ is, however, 
sufficient to explain why this should be so. 

To illustrate, let us consider the following. The sine of an angle is 
between 0 and 0.5 whenever the angle is between 0 and $\pi/6$. It is 
between 0.5 and 1.0 when the angle is between $\pi/6$ and $\pi/2$. Since 
the second range of angles is twice the first yet the range of the values of 
the sine is the same, it follows that if all angles in the range 0 to $\pi$ 
are selected equally often there will be twice as many sine values seen 
above 0.5 as below. 

When allowance is made for the addition of noise, the production of a 
distribution such as that of Fig.~\ref{fig7} for the average when angles are sampled 
uniformly comes as no surprise. Clearly, as the experimenters themselves 
recognise, the dip is not in itself sufficient to prove the non-classical 
state of the light. For this, direct measurement of the Wigner density is 
required, but there is a problem here. No actual direct measurement is 
possible, so it has to be estimated, and the method proposed is the Radon 
transformation \cite{Leonhardt:1997}. It is claimed \cite{Grangier:2005} 
that in \textit{other} experiments Wigner densities calculated by this procedure have 
shown negative regions, but perhaps the method should be checked for 
validity?

In any event, as already explained, the natural hidden variable relevant to 
the proposed experiment is the phase of the individual pulse, not any 
statistical property of the whole ensemble.

\section{Suggestions for extending the experiment}
The basic set-up would seem to present an ideal opportunity for investigation of some key 
aspects of the quantum and classical models, as well as the operation of the 
Bell test ``detection loophole''. 

\textbf{The operation of the detection loophole} could be illustrated if, 
instead of using the digitised difference voltages of the homodyne 
detectors, the two separate voltages are passed through discriminators. The 
latter operate by applying a threshold voltage that can be set by the 
experimenter and counting those pulses that exceed it. These can be used in 
a conventional CHSH Bell test, i.e.~using total observed coincidence count as 
denominator in the estimated quantum correlations $E$. 
The model that has been known since Pearle's time (1970) predicts that, as 
the threshold voltage used in the discriminators is increased and hence the 
number of registered events decreased (interpreted in quantum theory as the 
detection efficiency being decreased), the CHSH test statistic $S$, if calculated 
using estimates $E = (N_{++} + 
N_{--} - N_{+-} - N_{-+}) / (N_{++} + N_{--} + N_{+-} + N_{-+})$, will increase. 
If noise levels are low, it may well exceed the Bell limit of 2.

\textbf{The existence of the two phase sets} could also be investigated if 
either the raw voltages or the undigitised difference voltages are analysed. 
So long as the noise level is low, the existence of the two superposed 
curves, one for $\alpha = 0$ and the other for $\alpha =\pi$, should 
be apparent.  It would be interesting to investigate how the pattern changed
as optical path lengths were varied.  Schiller's pattern might be hard to 
reproduce using long path lengths, where exact equality is needed unless the
light is monochromatic.

If the primary goal of the experimenter is clearly set out to be the 
comparison of the performance of the two rival models, rather than merely 
the conduct of a Bell test, further ideas for modifying the set-up will 
doubtless emerge when the first experiments have been done.

\section{Conclusion}
The proposed experiments would, \textit{if the ``non-classicality'' of the 
light could be demonstrated satisfactorily}, provide a definite answer one way or the 
other regarding the reality of quantum entanglement. They could usefully be 
extended to include empirical investigations into the operation of the Bell 
test detection loophole. Perhaps more importantly, though, they present 
valuable opportunities to compare the performance of the two theories in 
both their total predictive power and their comprehensibility. Are 
parameters such as ``Wigner density'' and ``degree of squeezing'' really the 
relevant ones, or would we gain more insight into the situation by talking 
only of frequencies, phases and intensities? Parameters such as the 
detection efficiency and the transmittance of the beamsplitters will 
undoubtedly affect the results, but do they do this in the way the quantum 
theoretical model suggests? It will take considerably more than just the 
minimum runs needed for the Bell test if we are to find the answers.

The detailed predictions of the local realist model cannot be given until 
the full facts of the experimental set-up and the performance of the various 
parts are known, but it gives, in any event, a simple explanation of the 
double-peaked nature of the distribution of voltage differences. The peaks 
arise naturally from the way in which homodyne detection works, and the 
quantum theoretical idea that they are one of the indications of a non-classical beam 
or of negative Wigner density would not appear to be justifiable. The idea 
that a classical beam can become non-classical by the act of ``subtracting 
a photon'' is, equally, of doubtful validity. In the classical 
model, the only effect of the subtraction and detection of part of each beam 
is to aid the selection for coincidence analysis of those pulses that are 
likely to be most strongly correlated.

\section{Acknowledgements}
I am grateful to Ph.~Grangier for drawing my attention to his team's 
proposed experiment \cite{Sanchez:2004}, and to him and A.~I.~Lvovsky for helpful discussions.

\appendix
\section{Alternative classical derivation of the homodyne detection 
formula}
A derivation is given here that does not involve complex numbers and hence 
confirms that the equations in the text are ordinary wave equations, not 
quantum-mechanical ``wave functions''. 

The relationship between intensity 
difference and the local oscillator phase can be checked as follows:

\bigskip
\noindent Assume the two input beams are
\medskip

\par Experimental beam: $E \cos \phi$ , where, as before, $\phi =\omega t$
\nopagebreak 
\par Local oscillator: $E_{L} \cos (\phi +\theta )$

\medskip
\noindent Then the output beams, assuming a 50-50 beamsplitter and no losses, can be 
written
\medskip

Reflected beam: 
\begin{eqnarray}
\label{eqA1}
E_r & = & \frac{1}{\sqrt 2} (E\cos (\phi + \pi /2) + E_L \cos (\phi + \theta))
\nonumber \\
    & = & \frac{1}{\sqrt 2} (-E\sin \phi + E_L \cos (\phi + \theta))
\end{eqnarray}

Transmitted beam: 
\begin{eqnarray}
\label{eqA2}
E_t  & = & \frac{1}{\sqrt 2} (E\cos \phi + E_L \cos (\phi + \theta + \pi /2))
\nonumber \\
     & = & \frac{1}{\sqrt 2} (E\cos \phi - E_L \sin (\phi + \theta)).
\end{eqnarray}

\noindent Let us define a (constant) angle $\psi$ such that $\tan \psi = E / E_{L}$,
 making $E \cos \psi = E_{L } \sin \psi$.

\medskip
\noindent
Consider the case when $\theta = 0$.  We have

\begin{eqnarray}
\label{eqA3}
E_r & = & \frac{1}{\sqrt 2} (E/\sin \psi )(-\sin \psi \sin \phi +\cos \psi \cos \phi )
\nonumber \\
    & = & \frac{1}{\sqrt 2} (E/\sin \psi )\cos (\psi +\phi ),
\end{eqnarray}
\noindent
so that amplitude is proportional to $\frac{1}{\sqrt 2} E / \sin \psi$ and 
intensity to $\frac{1}{2} E^{2} / \sin^{2} \psi$.

Similarly, the intensity of the transmitted beam is also proportional to 
$\frac{1}{2} E^{2} / \sin^{2} \psi$.  The voltage difference from the homodyne 
detector is therefore expected to be zero. A zero difference is also 
found for $\theta =\pi $, but other values of $\theta $ produce more 
interesting results.
\medskip 

\noindent For example, for $\theta =\pi /2$ we find
\begin{eqnarray}
\label{eqA4}
E_r & = & \frac{1}{\sqrt 2} (E/\sin \psi )(-\sin \psi \sin \phi - \cos \psi \sin \phi )
\nonumber \\
    & = & \frac{1}{\sqrt 2} (E/\sin \psi )\sin \phi (-\sin \psi -\cos \psi ),
\end{eqnarray}
\begin{eqnarray}
\label{eqA5}
E_t & = & \frac{1}{\sqrt 2} (E/\sin \psi )(\sin \psi \cos \phi -\cos \psi \cos \phi )
\nonumber \\
    & = & \frac{1}{\sqrt 2} (E/\sin \psi )\cos \phi (\sin \psi -\cos \psi ).
\end{eqnarray}

\noindent The difference in intensities is therefore proportional to 
\begin{eqnarray}
\label{eqA6}
\mbox{Difference} & = & \frac{1}{2} (E^2/\sin ^2\psi )[(-\sin \psi - \cos \psi )^2
\nonumber \\
       & - & (\sin \psi - \cos \psi )^2]
\nonumber \\
       & = & \frac{1}{2} (E^2/\sin ^2\psi )4\sin \psi \cos \psi )
\nonumber \\
       & = & \frac{1}{2} E^2\cos \psi /\sin \psi 
\nonumber \\
       & = & 2E^2E_L /E
\nonumber \\
       & = & 2EE_L .
\end{eqnarray}

This is consistent with the result obtained by the method in the main text, 
so it seems safe to accept that as being correct.


\begin{thebibliography}{21}
\bibitem{prediction} The predicted violation is in any case small, so failure may be put 
down to other ``experimental imperfections''.
\bibitem{phase offset} The phase offset depends on the difference in optical path lengths, 
which will not in practice be exactly constant due to thermal oscillations. If a 
complete model is ever constructed, this should therefore be a parameter.
\bibitem{equal probability} It is possible that there will not be quite perfect symmetry 
between the two values of $\alpha$, if the frequency doubling process does not 
quite eliminate the initial laser frequency.
\bibitem{quantum correlation} In the derivation of Bell inequalities such as the CHSH 
inequality, the statistic required is simply the expectation value of the product of the 
outcomes.  This is commonly referred to in this context as the ``quantum correlation''. 
Only when there are no null outcomes, possible outcomes being restricted to just $+1$ 
or $-1$, does it coincide with ordinary statistical correlation.
\end{thebibliography}
\end{document}